\documentclass[sigconf,preprint]{acmart}
\graphicspath{{resources/}} 

\usepackage{subcaption}

\newcommand{\ptitle}[1]{\emph{#1}.}

\DeclareMathOperator*{\argmax}{arg\,max}

\AtBeginDocument{%
  \providecommand\BibTeX{{%
    \normalfont B\kern-0.5em{\scshape i\kern-0.25em b}\kern-0.8em\TeX}}}

\setcopyright{none}
\renewcommand\footnotetextcopyrightpermission[1]{}
\begin{document}

\title{CSNE: Conditional Signed Network Embedding}

\author{Alexandru Mara}
\email{alexandru.mara@ugent.be}
%
\author{Yoosof Mashayekhi}
\email{yoosof.mashayekhi@ugent.be}
\affiliation{%
  \institution{Ghent University}
  \city{Ghent}
  \postcode{9052}
  \country{Belgium}
}

\author{Jefrey Lijffijt}
\email{jefrey.lijffijt@ugent.be}
\author{Tijl De Bie}
\email{tijl.debie@ugent.be}
\affiliation{%
  \institution{Ghent University}
  \city{Ghent}
  \postcode{9052}
  \country{Belgium}
}

\renewcommand{\shortauthors}{Mara and Mashayekhi, et al.}

\begin{abstract}

Signed networks are mathematical structures that encode positive and negative relations between entities such as friend/foe or trust/distrust. Recently, several papers studied the construction of useful low-dimensional representations (embeddings) of these networks for the prediction of missing relations or signs. Existing embedding methods for sign prediction generally enforce different notions of status or balance theories in their optimization function. These theories, however, are often inaccurate or incomplete, which negatively impacts method performance. 

In this context, we introduce conditional signed network embedding (CSNE). Our probabilistic approach models structural information about the signs in the network separately from fine-grained detail. Structural information is represented in the form of a prior, while the embedding itself is used for capturing fine-grained information. These components are then integrated in a rigorous manner. CSNE's accuracy depends on the existence of sufficiently powerful structural priors for modelling signed networks, currently unavailable in the literature. Thus, as a second main contribution, which we find to be highly valuable in its own right, we also introduce a novel approach to construct priors based on the Maximum Entropy (MaxEnt) principle. These priors can model the \emph{polarity} of nodes (degree to which their links are positive) as well as signed \emph{triangle counts} (a measure of the degree structural balance holds to in a network).

Experiments on a variety of real-world networks confirm that CSNE outperforms the state-of-the-art on the task of sign prediction. Moreover, the MaxEnt priors on their own, while less accurate than full CSNE, achieve accuracies competitive with the state-of-the-art at very limited computational cost, thus providing an excellent runtime-accuracy trade-off in resource-constrained situations.


\end{abstract}



\maketitle
\newcommand{\bA}{\mathbf{A}}
\newcommand{\bB}{\mathbf{B}}
\newcommand{\bC}{\mathbf{C}}
\newcommand{\bX}{\mathbf{X}}
\newcommand{\bx}{\mathbf{x}}
\newcommand{\bW}{\mathbf{W}}
\newcommand{\bw}{\mathbf{w}}
\newcommand{\bs}{\mathbf{s}}
\newcommand{\bI}{\mathbf{I}}
\newcommand{\bL}{\mathbf{L}}
\newcommand{\bD}{\mathbf{D}}
\newcommand{\bM}{\mathbf{M}}
\newcommand{\bMu}{\mathbf{\mu}}
\newcommand{\bN}{\mathbf{N}}
\newcommand{\bPhi}{\mathbf{\Phi}}
\newcommand{\bSigma}{\mathbf{\Sigma}}
\newcommand{\bLLambda}{\mathbf{\Lambda}}
\newcommand{\bPi}{\mathbf{\Pi}}
\newcommand{\bQ}{\mathbf{Q}}
\newcommand{\res}{\Delta\bOne}
\newcommand{\bDelta}{\Delta}
\newcommand{\bZero}{\mathbf{0}}
\newcommand{\bz}{\mathbf{z}}
\newcommand{\bZ}{\mathbf{Z}}
\newcommand{\bOne}{\mathbf{1}}
\newcommand{\bLambda}{\mathbf{\lambda}}

\newcommand{\hbX}{\hat{\bX}}
\newcommand{\hbx}{\hat{\bx}}
\newcommand{\hbW}{\hat{\bW}}
\newcommand{\hbw}{\hat{\bw}}
\newcommand{\hbPi}{\hat{\bPi}}

\newcommand{\bbR}{\mathbb{R}}
\newcommand{\bbD}{\mathbb{D}}
\newcommand{\bbE}{\mathbb{E}}

\newcommand{\cP}{\mathcal{P}}
\newcommand{\cM}{\mathcal{M}}
\newcommand{\cN}{\mathcal{N}}
\newcommand{\cL}{\mathcal{L}}
\newcommand{\cO}{\mathcal{O}}

\newcommand{\Tr}{\text{Tr}}
\newcommand{\entropy}{-\int p_\bX\left(\bX\right)\log\left(p_\bX(\bX)\right)d\bX}

\section{Introduction}\label{sec_intro}
In recent years, signed networks have become prominent in online media representing friend/foe relations \cite{kunegis2009slashdot}, trust networks describing trust/distrust \cite{guha2004propagation}, natural language modelling for describing synonyms and antonyms \cite{islam2018signet}, and sentiment analysis denoting positive or negative opinions \cite{wang17shine}. Signed networks are powerful data representations that can describe complex interactions between heterogeneous entities represented by its nodes (or vertices). Generally, we refer to these interactions or connections between node-pairs as edges or links. In contrast to classical networks, where links simply denote a relation between entities, in signed networks these relations can be `positive' (e.g. friend, trust) or `negative' (e.g. foe, distrust).

An important task on signed networks is the prediction of signs between node-pairs for which this sign is unknown.
This contrasts with link prediction, where the purpose is to predict whether any given node-pair with unknown link status should be linked or not.
Sign prediction is an important problem in practice owing to the high cost of acquiring information on the signs in many domains.
In social networks, for example, interactions between users (e.g. messages exchanged, or connections made) can be efficiently tracked.
Unveiling the positive or negative nature of those relations between users (i.e. edge signs) from these interactions, however, is more challenging.
In a trust network it may be easy to keep track of who engaged in a transaction with whom,
but much less easy to understand the mutual trust relations underlying these transactions.
Note that sign prediction not only allows one to predict the sign of a link in the network,
it also allows one to predict the sign of a link the existence of which is not established.
In social networks, for example, this would allow one to assess whether two people are likely to become friends or foes if introduced to each other.
Sign prediction has thus been the topic of many recent studies \cite{yuan2017sne, wang2017signed, kim2018side, song2018learning}.

A particular class of approaches extensively used for the analysis of \emph{unsigned} networks consists of representation learning methods, also called network embedding methods \cite{perozzi2014deepwalk, cne2019}. These techniques model nodes as low-dimensional vectors in ${\rm I\!R}^d$. The underlying idea is that similar nodes in the graph are mapped to close-by vectors in the embedding space. Using these representations, traditional machine learning methods can be applied on network data to perform downstream tasks such as link prediction \cite{grover2016node2vec, wang2018shine}, information diffusion \cite{gao2017novel, li2017deepcas, bourigault2016representation}, and multi-label classification \cite{tang2015line}. 

Despite their success, however,
unsigned network embedding techniques are not directly applicable to signed networks.
Indeed, in signed networks node-pairs can have three possible states:
unlinked, positively linked, and negatively linked.
Unsigned network embedding methods only consider two possible states (unlinked and linked).
Thus, to apply them unaltered to signed networks,
one would have to ignore the distinction between two of the three states,
e.g. ignoring the difference between negatively linked and unlinked.
This is problematic, as the semantics of these states are often very different.
For example, unlinked node-pairs are often to be interpreted as having an \emph{unknown} sign.
%
%
Moreover, for many signed networks it may be useful to account for balance theory \cite{heider46balanceth} in the model,
which is conceptually impossible without distinguishing these three possible states of node-pairs.
Balance theory suggests that triads with an odd number of negative connections are unstable and, thus, less likely to form or persist over time.
For example, in a trust network structural balance means one is less likely to trust someone who is distrusted by a friend. 
 
To overcome the aforementioned difficulties, several embedding methods for signed networks have been proposed, such as SiNE \cite{wang2017signed}, Signet \cite{islam2018signet} and SNE \cite{yuan2017sne}. These methods adopt random walk or probabilistic strategies to learn the structure of signed networks and have been shown to outperform unsigned embedding methods for downstream tasks, such as sign prediction. Nevertheless, these methods present two major drawbacks. On the one hand, they often impose constraints on the learning process that \emph{strictly} enforce status or balance theories, even though these theories often holds to only a certain extent \cite{Leskovec10}. On the other hand, they often suffer from overfitting as only the connected node-pairs are used in the embedding learning process. 

To address these shortcomings, we propose a novel probabilistic approach which we coin Conditional Signed Network Embedding (CSNE).
The idea of CSNE is to find an embedding for the nodes in the network that is maximally informative of the signs in the network
\emph{conditioned} on a probabilistic prior for these signs.
These two distributions are then combined in a rigorous manner.
The rationale is the observation that embeddings are often unsuited for representing complex structural information, 
while they are strong at representing fine-grained local information \cite{Lu18Struc}.
If we are thus able to use a probabilistic prior to represent structural information about the network,
both types of information are represented and can be leveraged for sign prediction.


More specifically, inspired by Conditional Network Embedding (CNE) \cite{cne2019},
we construct a likelihood function for the embedding (i.e. a distribution for the signs conditional on the embeddings),
and infer the embedding using the Maximum Likelihood principle.
The key idea of CSNE is to construct this likelihood function using Bayes rule from a prior distribution for the signs,
and a conditional distribution for the embedding conditioned on the signs.
This approach allows us to conveniently model structural information in a suitable prior while the embedding itself can focus exclusively on learning fine-grained local connectivity relations between nodes
that are more easily expressed in a metric space.

CSNE could work, at least in principle, with any prior for the signs.
Yet, in this paper we propose a novel approach for inferring such priors. Our approach can model two kinds of structural information that existing methods struggle to represent, but that may be important in signed networks.
The first is that within a network, some nodes may have a more positive (or negative) inclination than others. When modelling, for instance, social interactions, certain users will be liked more often, while others are disliked more often depending on their personality, ideology etc. For new connections formed by these users, we may expect such tendency to persist. We refer to this underlying tendency as the \emph{polarity} of a node.
The second is that structural balance rarely holds exactly, but it may hold to a certain extent.
The extent to which it holds can be quantified in terms of the number of balanced triangles (where structural balance holds)
versus the number of unbalanced triangles (where it does not hold) in the network.
For example in a trust network, if the rate at which trust is being breached is high, structural balance will hold less strongly.
Indeed, even if in an equilibrium situation structural balance would hold, occasional breach of trust will disrupt this balance.

Unfortunately, no approaches have been proposed in the literature modelling such information, for use as a prior in CSNE.
Thus, building upon recent work from \citet{adriaens2020scalable} on efficient Maximum Entropy (MaxEnt) models for (unsigned) graphs,
we show how both polarity and the degree to which structural balance holds can be modelled in an accurate and highly scalable manner.
This approach is a second main result in this paper.

The \textbf{main contributions} of this paper are:
\begin{itemize}
    \item We propose a new probabilistic method for signed network embedding and sign prediction which we coin Conditional Signed Network Embedding (CSNE). Our method computes the most informative embeddings w.r.t. the proposed prior on the network structure. 
	\item We design a new edge-independent MaxEnt prior for signed networks. This prior models two structural properties of signed networks: the proportion of positive to negative connections of each node or node \textit{polarity} and the number of balanced and unbalanced triangles formed by each edge or \textit{triangle counts}.
    \item We provide extensive experiments, comparing with existing methods for signed network embedding and sign prediction.
\end{itemize}

The \textbf{benefits of these contributions} are as follows:
\begin{itemize}
\item CSNE is a \emph{conceptually novel} and \emph{mathematically rigorous} probabilistic approach to signed network embedding,
capable of modelling both structural and fine-grained information.
Its \emph{modular design} means that future research developments on priors for signed networks have the potential to further boost CSNE's performance.
\item The proposed MaxEnt prior, which is \emph{conceptually simple and intuitive}, is \emph{extremely efficient} to compute
(an order of magnitude faster than SIGNet, the fastest state-of-the-art method for sign prediction).
When used on its own for sign prediction, it already \emph{uniformly outperforms} all state-of-the-art baselines
evaluated across a diversity of commonly used benchmark networks.
\item \emph{CSNE} with the proposed MaxEnt prior \emph{further improves upon this accuracy} for sign prediction,
while requiring computation times equal to or only slightly higher than the fastest state-of-the-art baseline method (SIGNet).
\end{itemize}

The remainder of the paper is organized as follows. In Sec.~\ref{sec_related}, we briefly discuss the related work. In Sec.~\ref{sec_proposed}, we introduce our proposed method. For ease of exposition, we first discuss the MaxEnt prior, before discussing CSNE and how the prior can be used as a building block in CSNE. Our experimental setup is described in Sec.~\ref{sec_exp}. We report the results in Sec.~\ref{sec_results}. Finally, Sec.~\ref{sec_conclusion} concludes this paper and summarizes open questions.

\section{Related Work}\label{sec_related}

Our paper is primarily related to the analysis of signed networks and the prediction of missing signs \cite{yuan2017sne, wang2017signed, kim2018side, song2018learning}. Our modular approach, additionally, incorporates ideas from two related fields.

First, CSNE connects to a large body of research in the field of unsigned network representation learning. Early approaches such as Laplacian Eigenmaps \cite{belkin2002laplacian} and Locally Linear Embeddings \cite{Roweis2000lle} were motivated by dimensionality reduction. More recently, embedding methods have been used to bridge the gap between traditional machine learning and network structured data \cite{liben2007link, bhagat2011node, papadopoulos2012community}. Methods such as DeepWalk \cite{perozzi2014deepwalk} and Node2vec \cite{grover2016node2vec} have been proposed to learn embeddings by leveraging a random walk strategy on the graphs. Other approaches such as GraRep \cite{cao2015grarep} and Arope \cite{Zhang2018arope} aim to model high order proximities between nodes in the networks. A recent probabilistic approach, CNE, can efficiently incorporate prior information in the embedding learning process. This method has also been shown in a recent empirical study by \citet{mara2020network} to largely outperform other embedding methods for the task of link prediction. For these reasons, our research incorporates ideas introduced in CNE to the context of signed network embedding. As mentioned earlier, however, the aforementioned methods are not directly applicable to the analysis of signed network. 

Second, the proposed probabilistic approach for modelling structural properties of signed networks
borrows ideas from the field of maximum entropy random graph models \cite{Jaynes57maxent, ParkNewman, DuijnERG},
and most directly from \citet{DeBie2011}.
Our approach additionally incorporates recent results from \citet{adriaens2020scalable} to address a fundamental challenge of these models:
the difficulty of modelling structural properties, which tend to introduce dependencies between the variables involved,
without loosing computational tractability.

Several methods for the specific task of signed network embedding have been proposed. One of the first approaches in this field is SNE \cite{yuan2017sne}, which computes embeddings through a log-bilinear model that incorporates link types in a given path in the network. More recent approaches \cite{wang2017signed, islam2018signet, lu2019ssne}, impose constraints in the optimization process to enforce different notions of structural balance. SiNE \cite{wang2017signed} proposes a deep learning framework for signed network embedding and incorporates the extended structural balance theory proposed in \citet{cygan2012sitting}. SIGNet \cite{islam2018signet} is a scalable node embedding method for signed networks. Akin to CSNE, this method aims to incorporate the notion of balance theory introduced by \citet{heider46balanceth}. Targeted node sampling \cite{mikolov2013distributed}, which extends negative sampling techniques from classical word2vec algorithms, is used to maintain structural balance in higher order neighbourhoods. And in \citet{lu2019ssne}, the authors propose SSNE, a method for embedding directed signed networks that considers status theory \citep{guha2004propagation, leskovec2010signed} in its learning phase. The status of a node is determined based on positive and negative links and a ranking is computed. 
In contrast to these methods, CSNE does not assume a notion of balance must hold exactly.
Instead, our method learns the extent to which balance holds for every network.
Finally, in another recent work, two methods nSNE and lSNE were proposed for learning node and edge embeddings in signed networks \cite{song2018learning}.

\section{Proposed Method}\label{sec_proposed}
In this section we introduce CSNE. We start in Sec.~\ref{ssec_background} by introducing basic concepts and notation for signed networks and sign prediction. In Sec.~\ref{ssec_maxent}, we recap some prior research on MaxEnt graph models and introduce a novel edge-independent MaxEnt distribution that allows one to efficiently model polarity and different types of triangle counts (to model the extent to which structural balance holds) in signed networks. Finally, in Sec.~\ref{ssec_csne} we describe the overall CSNE method, and how the MaxEnt model can be used as a building block thereof. In what follows, we limit our analysis exclusively to undirected networks. 

\subsection{Concepts and notation\label{ssec_background}}

\subsubsection{Signed Networks}
We represent an undirected signed network by $G=(V, E)$ with $|V|=n$ nodes and $|E|=m$ edges where $E \subseteq {V \choose 2}$. Each node $i \in V$ denotes an entity and edges $\{i,j\} \in E$ represent unordered relations between two entities $i$ and $j$. Edges $\{i,j\}$ can describe positive or negative relations in signed networks. A function $s: E \rightarrow \{-1, +1\}$ is used to map edges to their respective signs.
We denote the set of positive links i.e. $\{\{i,j\} \in E \mid s(\{i,j\})=1\}$ by $E^+$, and negative links $\{\{i,j\} \in E \mid s(\{i,j\})=-1\}$ by $E^-$.
The adjacency matrix of a signed network $G$ is represented as $\mathbf{\hat{A}}\in\mathcal{A}=\{-1,0,1\}^{n\times n}$ with entries $\hat{a}_{ij} \in \{-1,0,1\}$, and with $\hat{a}_{ij}=0$ if $\{i,j\}\not\in E$ and $\hat{a}_{ij}=s(\{i,j\})$ otherwise.%
\footnote{In some settings, it might be useful to consider a fourth possible state of a node-pair $\{i,j\}\in {V \choose 2}$: linked (i.e. $\{i,j\}\in E$), but with unknown sign (i.e. $s(\{i,j\})$ unknown). Although semantically this is different, from a methodological perspective the distinction between such node-pairs and unlinked node-pairs is irrelevant, as the only node-pairs used for training in any existing method (including ours) are those for which the sign is known.}

\subsubsection{Sign Prediction\label{ssec_bgsp}}
This task amounts to, given an observed signed network $G=(V, E)$, inferring the signs $s(\{i,j\}) \in \{-1,1\}$ of unobserved node-pairs $\{i,j\}\in{{V \choose 2}\setminus E}$.

\subsection{A MaxEnt distribution for edge signs\label{ssec_maxent}}

Maximum entropy random graph models are statistical models used for network inference based on the \textit{principle of maximum entropy} \cite{Jaynes57maxent}. This principle states that the best estimate of a distribution given certain constraints is the one with highest entropy amongst those satisfying said constraints. In practice, these constraints are often derived from empirical data.

In the context of \textit{unsigned} networks, MaxEnt modelling works as follows. Given an observed \emph{binary} adjacency matrix $\hat{\mathbf{A}} \in \mathcal{A}_{bin}= \{0,1\}^{n\times n}$ and a set of statistics (e.g. node degrees, assortativity, densities of particular blocks, etc.) considered characteristic for this network, 
one seeks the MaxEnt distribution $P$ over random $\mathbf{A} \in \mathcal{A}_{bin}$ such that the expected values of these statistics are equal to the values empirically observed on $\hat{\mathbf{A}}$. For instance, one can seek the MaxEnt distribution $P$ which preserves the node degrees of $\hat{\mathbf{A}}$ i.e. $\mathbb{E}_P[\sum_i \mathbf{A}_{i,j}] = \sum_i \hat{\mathbf{A}}_{i,j}$.


Here, our objective is to learn a MaxEnt distribution $P(\textbf{A})$ for the \emph{signs} of all node-pairs $\{i,j\}\in{V \choose 2}$. 
We use a similar formulation to that of MaxEnt models for unsigned graphs, with a key difference:
the random variables are now the signs $s(\{i,j\})$ instead of whether a node-pairs is connected or not.
In summary, we thus aim to find a distribution over the set $\mathcal{A}_{sign}= \{-1,1\}^{n\times n}$ of all possible sign matrices $\mathbf{A}$ of size $n$,
with expected values for certain important statistics equal to their values on the empirically observed signs.
Of course, as the signs are known only for the linked node-pairs, the statistics are computed based on the linked node-pairs only.

As observed by various authors \cite{Goodreau2007AdvancesIE, adriaens2020scalable}, two major drawbacks when fitting MaxEnt models are scalability and generalization to arbitrary constraints.
Thus, a key challenge in the development of MaxEnt models is to identify statistics that are characteristic for the data,
while using them as constraints is computationally tractable.
We already argued that node polarity statistics, and statistics on the number of balanced/unbalanced triangles, are both useful to use as constraints.
Now, we will show that their use as constraints in MaxEnt modelling is computationally tractable as well.
Our approach is based on recent work by \citet{adriaens2020scalable}, which identifies a broad class of statistics that can be used to formulate these constraints.

Let $\mathbf{F}\in\mathbb{R}^{n\times n}$ denote a real-valued matrix with $f_{ij}$ the element on row $i$ and column $j$, referred to as a \emph{feature matrix}.
Rephrased for our current context, \citet{adriaens2020scalable} then demonstrate that constraints on statistics of the form
\begin{align}\label{eq:Fstatistic}
\gamma(\mathbf{A})\triangleq\sum_{\{i,j\} \in E} f_{ij}a_{ij}
\end{align}
can be used efficiently for MaxEnt modelling.
Note that such statistics satisfy our requirement that pertain to the edges only, such that their empirical values can be computed.
They can also be used to calculate the polarity of each node, as well as the number of triangles with particular sign patterns, as discussed next.

\paragraph{Polarity.} We formally define the polarity of a node as the sum of the signs of the edges incident to this node.
To model the polarity of a node $l\in V$, one can use a statistic based on the feature matrix $\mathbf{F}_l$ defined as follows:
\begin{equation}
f_{ij}^{l} = 
\begin{cases}
1, & \text{iff $i=l$}\\
0, & \text{otherwise}
\end{cases}
\end{equation}
Indeed, for such a feature matrix, $\sum_{\{i,j\} \in E} f_{ij}a_{ij}=\sum_{j: \{l,j\}\in E} a_{lj}$, equal to the polarity of node $l$ as required.
Let us denote the corresponding polarity statistic for node $l$ as $\gamma_l$.

\paragraph{Counting triangles with various sign patterns}
In addition to the polarity statistics (one per node),
we also use three statistics that jointly describe how many triangles the network contains with three positive signs (+++ triangles), with two positive signs (++- triangles), with one positive sign (+-{}- triangles), and with all negative signs (-{}-{}- triangles).
Note that the total number of triangles is fixed, so three of these constraints will suffice.
Moreover, using any linear combination of these statistics would be equivalent.
In particular, it is mathematically convenient to count the following three statistics which jointly can be used:
\begin{itemize} 
\item {\bf Triangle statistic $\gamma_{++}$} This statistic is defined as the number of +++ triangles minus the number of ++- triangles. This can be counted in the form of Eq.~(\ref{eq:Fstatistic}) by defining $f_{ij}$ as the number wedges connecting $i$ and $j$ with positive edges only, or formally:
\begin{equation}
f_{ij}^{++} = \sum_{k:\{k,i\}\in E\land\{k,j\}\in E} \frac{\hat{a}_{ik} +1}{2} \frac{\hat{a}_{kj} +1}{2}.
\end{equation}
Then $\sum_{\{i,j\} \in E} f_{ij}^{++}\hat{a}_{ij} = \sum_{\{i,j\}\in E^+} f^{++}_{ij} - \sum_{\{i,j\}\in E^-} f^{++}_{ij}$,
which is indeed equal to the number of +++ triangles minus the number of ++- triangles.

\item {\bf Triangle statistic $\gamma_{+-}$} This statistic is defined as the number of ++- triangles minus the number of +-{}- triangles. It can be computed as in Eq.~(\ref{eq:Fstatistic}) by defining the $f_{ij}^{+-}$ as the number of wedges with differing signs that connect $i$ and $j$, or formally:
\begin{equation}
f_{ij}^{+-} = \sum_{k:\{k,i\}\in E\land\{k,j\}\in E} \frac{1-\hat{a}_{ik}\hat{a}_{kj}}{2}.
\end{equation}

\item {\bf Triangle statistic $\gamma_{--}$} This statistic is defined as the number of +-{}- triangles minus the number of -{}-{}- triangles. It can be computed as in Eq.~(\ref{eq:Fstatistic}) by defining the $f_{ij}^{--}$ as the number of wedges with two minus signs that connect $i$ and $j$, or formally:
\begin{equation}
f_{ij}^{--} = \sum_{k:\{k,i\}\in E\land\{k,j\}\in E} \frac{1-\hat{a}_{ik}}{2} \frac{1-\hat{a}_{kj}}{2} .
\end{equation}
\end{itemize}
Note that these three statistics jointly, together with the fixed overall triangle count,
indeed uniquely define the number of triangles with any sign pattern.
For example, denoting the total number of triangles as $t$,
it is easy to verify that the number of ++- triangles can be computed as $\frac{t+\gamma_{--}+2\gamma_{+-}-\gamma_{++}}{4}$.

Thus the MaxEnt distribution can be found by solving the following convex optimization problem:
\begin{align}\label{eq:maxentGeneral}
	\displaystyle &\argmax_{P(\textbf{A})} \quad -\displaystyle \mathbb{E}_P[\log P(\textbf{A})], \\
	\text{s.t. }\quad &\mathbb{E}_{\mathbf{A}~P}[\gamma_l(\mathbf{A})] = c_l  \quad \forall l=1,\ldots,n,\notag\\
	&\mathbb{E}_{\mathbf{A}~P}[\gamma_{++}(\mathbf{A})] = c_{++},\notag\\
	&\mathbb{E}_{\mathbf{A}~P}[\gamma_{+-}(\mathbf{A})] = c_{+-},\notag\\
	&\mathbb{E}_{\mathbf{A}~P}[\gamma_{--}(\mathbf{A})] = c_{--},\notag
\end{align}
where $c_l=\gamma_l(\hat{\mathbf{A}})$, $c_{++}=\gamma_{++}(\hat{\mathbf{A}})$, $c_{+-}=\gamma_{+-}(\hat{\mathbf{A}})$, and $c_{--}=\gamma_{--}(\hat{\mathbf{A}})$ denote the empirically observed values of the statistics.

As shown by \citet{adriaens2020scalable}, the formulation in Eq.~(\ref{eq:maxentGeneral}) factorizes as a product of independent Bernoulli distributions.
The solution, in this case, is of the following form:
\begin{align}\label{eq_prior}
	P(\mathbf{A}) = \prod_{\{i,j\} \in E} P(a_{ij}=1)^{\frac{\hat{a}_{ij}+1}{2}}(1-P(a_{ij}=1))^{1-\frac{\hat{a}_{ij}+1}{2}}.
\end{align}
Moreover, the `success probabilities' $P(a_{ij}=1)$ for node-pairs $\{i,j\}$ are equal to:
\begin{equation}\label{eq_sucprob}
P(a_{ij} = 1) = \dfrac{\exp(\textstyle\sum_{l=1}^n f_{ij}^l \lambda_l + f_{ij}^{++} \lambda_{++} + f_{ij}^{+-} \lambda_{+-} + f_{ij}^{--} \lambda_{--})}{1+\exp(\textstyle\sum_{l=1}^n f_{ij}^l \lambda_l + f_{ij}^{++} \lambda_{++} + f_{ij}^{+-} \lambda_{+-} + f_{ij}^{--} \lambda_{--})},
\end{equation}
where $\lambda_l, \lambda_{++}, \lambda_{+-}$, and $\lambda_{--}$ denote the Lagrange multipliers associated with the respective constraints in Eq.~(\ref{eq:maxentGeneral}).
The optimal values of these Lagrange multipliers can be found by unconstrained minimization of the convex Lagrange dual function:
\begin{align}\label{eq:lagrangian}
&L(\lambda_1,\ldots,\lambda_M) =\\
&\textstyle\sum_{\{i,j\} \in E}\log(1+\exp(\textstyle\sum_{l=1}^n f_{ij}^l\lambda_l + f_{ij}^{++} \lambda_{++} + f_{ij}^{+-} \lambda_{+-} + f_{ij}^{--} \lambda_{--}))\nonumber\\
&-\sum_{l=1}^n c_l \lambda_l - c_{++} \lambda_{++} - c_{+-} \lambda_{+-} - c_{--} \lambda_{--}.\nonumber
\end{align}
Again, as shown by \citet{adriaens2020scalable}, this can be solved efficiently for very large networks by using equivalences between the Lagrange multipliers. In our proposed model, this reduces the number of free variable to $O(\sqrt{n})$. Finally, we observe that the MaxEnt prior encodes probabilities for each node-pair in the graph of being positive or negative. Therefore, this prior can be directly used for sign prediction.






\subsection{Conditional Signed Network Embedding\label{ssec_csne}}
As introduced in Sec.~\ref{sec_intro}, the MaxEnt model is required as a prior for CSNE. For ease of exposition, the prior was discussed in the previous section. In the current section, we introduce CSNE proper and explain how the MaxEnt prior is an integral component. However, before diving into the details of CSNE, we formally define the task of signed network embedding.

Embedding approaches learn a function $g: V \rightarrow {\rm I\!R}^d$ which maps nodes in the network to d-dimensional real-valued vectors. These representations are generally denoted as $\mathbf{X} = (\mathbf{x}_1, \mathbf{x}_2, \dots, \mathbf{x}_m)' \in {\rm I\!R}^{m \times d}$, where $\mathbf{x}_i$ is the embedding corresponding to node $i$. A common modelling assumption for NE methods is that similar nodes in the network must be mapped to close-by representations in the embedding space. For the particular case of signed network embedding, we additionally require the function $g: V \rightarrow {\rm I\!R}^d$ to map pairs $\{i,j\} \in E^-$ to more distant representations $\mathbf{x}_i$ and $\mathbf{x}_j$ in the embedding space, than pairs $\{i,j\} \in E^+$. At the same time, unlinked node-pairs are commonly not used in the embedding learning process i.e. their distance in unimportant.


CSNE aims to learn the most informative embedding $\mathbf{X}$ for a given signed network $G = (V, E)$ with adjacency matrix $\hat{\mathbf{A}}$. In simple terms, the objective is to find the embedding $\mathbf{X}$ that maximizes the likelihood of observing the signs on the edges in $\hat{\mathbf{A}}$. We formulate this optimization task as a Maximum Likelihood Estimation (MLE) problem, $argmax_{\mathbf{X}} P(\mathbf{A}|\mathbf{X})$. Akin to \citet{cne2019}, we do not postulate the likelihood function $P(\mathbf{A}|\mathbf{X})$ directly. Instead, we \textit{do} postulate the density function of the embedding $\mathbf{X}$ conditioned on the signed network, i.e. $p(\mathbf{X}|\mathbf{A})$. Then, we combine $p(\mathbf{X}|\mathbf{A})$ with the MaxEnt prior discussed in Sec.~\ref{ssec_maxent}, $P(\mathbf{A})$, by means of the Bayes formula. The likelihood function, thus, follows as $P(\mathbf{A}|\mathbf{X}) = \frac{p(\mathbf{X}|\mathbf{A})P(\mathbf{A})}{p(\mathbf{X})}$.
Independently modelling the density function of the embedding conditioned on the network and the MaxEnt distribution on the signs of $G$, has one major advantage. The prior, can encode properties of the network which do not have to be reflected by the learned embedding $\mathbf{X}$. Effectively, this means that CSNE can a make better use of the embedding space. 

The MaxEnt prior $P(\mathbf{A})$ has already been discussed in Sec.~\ref{ssec_maxent} and its expression is given in Eq.~\ref{eq_prior}.
%
%
For postulating the conditional density $p(\mathbf{X}|\mathbf{A})$, we only need to model the distances between the embeddings of node-pairs $\{i,j\}$. As metric, we use the Euclidean distance between the embeddings of the end nodes, i.e. $d_{ij} \triangleq ||\mathbf{x}_i - \mathbf{x}_j ||_2$. As already discussed, we require the distances between positively connected node-pairs $\{i,j\} \in E^+$ to be lower than those between negatively connected pairs $\{i,j\} \in E^-$. To model this, we use two half normal distributions with different spread parameters. Positively connected edges are generated from a half normal distribution with spread parameter $\sigma_1$ while negatively connected edges from a distribution with $\sigma_2$ where $\sigma_1 < \sigma_2$. Thus, we have:

\begin{align}
	p(d_{ij}\mid \{i,j\} \in E^+) &= \cN_+\left(d_{ij} | \sigma_1^2\right) \\
	p(d_{ij}\mid \{i,j\} \in E^-) &= \cN_+\left(d_{ij} | \sigma_2^2\right)
\end{align}

The conditional density $p(\mathbf{X}|\mathbf{A})$ can then be expressed as follows:

\begin{align} \label{eq:cond}
p(\mathbf{X}|\mathbf{A}) = \prod_{\{i,j\}\in E^+} \cN_+\left(d_{ij} | \sigma_1^2\right) \cdot \prod_{\{i,j\}\in E^-} \cN_+\left(d_{ij} | \sigma_2^2\right) 
\end{align}

The resulting likelihood function to optimize is:

\begin{align}\label{eq:loglik}
&P(\mathbf{A} | \mathbf{X}) = \frac{p(\bX | \mathbf{A}) P(\mathbf{A})}{p(\bX)} =  \frac{p(\bX | \mathbf{A}) P(\mathbf{A})}{\sum_{\mathbf{A}} p(\bX | \mathbf{A}) P(\mathbf{A})} \nonumber \\
& = \prod_{\{i,j\}\in E^+} \frac{\cN_+\left(d_{ij} |  \sigma_1^2\right)P(a_{ij}=1)}{\cN_+\left(d_{ij} |  \sigma_1^2\right)P(a_{ij}=1) + \cN_+\left(d_{ij} |  \sigma_2^2\right)(1-P(a_{ij}=1))} \nonumber\\
& \cdot \prod_{\{i,j\} \in E^-}\frac{\cN_+\left(d_{ij} |  \sigma_2^2\right)(1-P(a_{ij}=1))}{\cN_+\left(d_{ij} |  \sigma_1^2\right)P(a_{ij}=1) + \cN_+\left(d_{ij} |  \sigma_2^2\right)(1-P(a_{ij}=1))}
\end{align}

In order to maximize the likelihood function Eq.~(\ref{eq:loglik}) we use block stochastic gradient descent. The derivation of the gradient follows closely the one in \citet{cne2019}. Thus, we refer the user to this manuscript for more details. The gradient of the log-likelihood function in CSNE w.r.t. the embedding $\mathbf{x}_i$ of node $i$ reads:

\begin{align} \label{eq:dergrad}
\frac{\partial \ log(P(\mathbf{A} | \mathbf{X}))}{\partial \ \mathbf{x_i}} &= 2 \sum_{j:\{i,j\} \in E^+} (\mathbf{x}_i - \mathbf{x}_j)P(a_{ij}=-1|\mathbf{X}) \bigg(\frac{1}{\sigma_2^2} - \frac{1}{\sigma_1^2} \bigg) \nonumber \\
&+ 2 \sum_{j:\{i,j\} \in E^-} (\mathbf{x}_i - \mathbf{x}_j)P(a_{ij}=1|\mathbf{X}) \bigg(\frac{1}{\sigma_1^2} - \frac{1}{\sigma_2^2} \bigg)
\end{align}

Intuitively, the first summation in the gradient expression pulls the embedding of each node close to the embeddings of positively connected neighbours. The second summation, on the other hand, pushes this embedding far from those of negatively connected nodes. 


\section{Experimental Setup}\label{sec_exp}

To evaluate the performance of CSNE, we conducted experiments on six networks from four sources, which we introduce in Sec.~\ref{ssec_data}. We compared the empirical observations for CSNE with four recent methods for signed network embedding that are discussed in Sec.~\ref{ssec_methods}. Specifics of the test setup, including hyperparameters settings for all methods are outlined in Sec.~\ref{ssec_sp}. Reproducibility details are in Sec.~\ref{ssec_rep}.

\subsection{Datasets\label{ssec_data}}
We performed sign prediction evaluation using 6 different real-world datasets. The first two networks constitute snapshots from the Slashdot\footnote{\url{http://slashdot.org/}} \cite{leskovec2010signed} social news website, where users can create friends (positive links) and foes (negative links). These snapshots, Slashdot(a) and Slashdot(b), were obtained in November 2008 and February 2009, respectively. Epinions\footnote{\url{http://www.epinions.com/}} \cite{leskovec2010signed} is a product review website, where users can trust (positive links) or distrust (negative links) each other. Wiki-rfa \cite{west2014exploiting} contains votes of Wikipedia users endorsing or opposing candidates for adminship. Neutral votes, also present in the data, are not used. The dataset contains information on votes between 2003 and 2013. Lastly, we used two Bitcoin cryptocurrency trust networks, Bitcoin-$\alpha$\footnote{\url{http://www.btcalpha.com/}} and Bitcoin-otc\footnote{\url{https://www.bitcoin-otc.com/}}. These networks, obtained from \cite{kumar2016edge}, were gathered in order to identify transactions with fraudulent users. 

All these datasets are originally directed and we preprocessed the networks by removing the direction of links. For the experiments we have used the largest connected component in each dataset and ignore self-loops. The most relevant statistics of each network are summarized in Table~\ref{tab:networks}.

\begin{table}
  \centering
  \caption{Main statistics of the networks used for sign prediction evaluation.}
  \label{tab:networks}
  \scalebox{0.74}{%
  \begin{tabular}{lcccccc}
    \toprule
    Data & Slashdot(a) & Slashdot(b) & Epinions & Wiki-rfa & Bitcoin-$\alpha$ & Bitcoin-otc \\
    \midrule
    $|V|$ & 77350 & 82140 & 131828 & 11258 & 3783 & 5881 \\
    $|E|$ & 468554 & 500481 & 711210 & 171562 & 14124 &	21492 \\
    $|E|/|V|$ & 12.11 & 12.18 & 10.78 & 30.47 & 7.46 & 7.30 \\
    \% $|E^+|/|E|$ & 75\% & 76\% & 83\% & 77\% & 90\% & 85\% \\
    \% Bal. Tri. & 85\% & 86\% & 89\% & 73\% & 83\% & 85\% \\
    \bottomrule
  \end{tabular}
	}
\end{table}

\subsection{Comparison methods}\label{ssec_methods}

To be able to interpret the performance of CSNE in the context of existing methods, we performed the same sign prediction experiments using the following methods:

\begin{itemize}
  \item SiNE \cite{wang2017signed} uses a deep neural network architecture to learn node embeddings. The objective function optimized by SiNE satisfies the structural balance theory. In this method, nodes are expected to be closer to their friends than their foes. If no negative connections exist for specific nodes, virtual nodes and negative links to these are generated. 
  \item nSNE \cite{song2018learning}, similarly to SiNE, uses a deep neural network to learn node embeddings by leveraging second order proximities in the graph. At the same time, the method learns a mapping from node embeddings to edge embeddings.
  \item lSNE \cite{song2018learning} is a simplified version of nSNE where the mapping function from node embeddings to edge embeddings is assumed to be linear. The function can therefore be learned via gradient descent.
  \item SIGNet \cite{islam2018signet} uses a random walk strategy to determine node similarity on the graph and the Skip-Gram model to obtain node embeddings. The authors propose an extension to the negative sampling used in word2vec models to perform targeted node sampling and accommodate the main concepts of structural balance theory (i.e., balanced triangles are more likely than unbalanced triangles).
  
\end{itemize}


\subsection{Sign Prediction Test Setup\label{ssec_sp}}

\ptitle{General setup} As introduced in Sec.~\ref{ssec_bgsp}, sign prediction amounts to identifying the sign of unobserved connections between nodes in a given network. For performance evaluation of sign prediction, it is common to divide the given set of edges ($E$) into two disjoint subsets: the train edges ($E_{train}$) are used in the model learning phase, while the test edges ($E_{test}$) are used for assessing the prediction performance of the methods. The train and test sets are constructed such that $E_{test} \cup E_{train} = E$, and $E_{test} \cap E_{train}=\emptyset$. At training time, edges in $E_{test}$ are removed and the corresponding value in the adjacency matrix is set to $0$. At test time, the model is then evaluated on all $\{i,j\}\in E_{test}$ and predictions compared to $\hat{a}_{ij}$. In our evaluation, we selected 80\% of the total edges $\{i,j\} \in E$ for training ($E_{train}$) and the remaining 20\% for testing ($E_{test}$). Train edges were selected regardless of their sign, using the default sampling strategy of the \texttt{EvalNE} toolbox \citet{mara2019evalne}, which ensures the training network remains connected.
The sets $E_{train}$ and $E_{test}$ are expected to contain similar proportions of positive and negative edges as the original graph. Unless otherwise specified, all results reported are averages over three independent repetitions of the experiment with different train and test sets.

\ptitle{Edge embeddings} For prediction, nSNE, lSNE and our proposed methods can directly return the probability of an edge $\{i,j\}$ of being positive or negative. For SiNE and SIGNet, however, this is not the case. These methods only return node embeddings from which predictions must be derived. As shown by \citet{Gurukar2019NRL} an effective approach for obtaining predictions from node embeddings is by training a binary classifier on edge embeddings derived from the node embeddings. The embedding of a link $\{i,j\}$ can be computed by applying different operators $\circ$ to the embeddings of the incident nodes $i$ and $j$ i.e. $\mathbf{x}_{ij} = \mathbf{x}_i \circ \mathbf{x}_j$. In our evaluation, we selected the operators introduced in \citet{grover2016node2vec}, namely \textit{Average} ($(\mathbf{x}_i + \mathbf{x}_j)/2$), \textit{Hadamard} ($\mathbf{x}_i \cdot \mathbf{x}_j$), \textit{Weighted $L_1$} ($|\mathbf{x}_i - \mathbf{x}_j|$) and \textit{Weighted $L_2$} ($|\mathbf{x}_i - \mathbf{x}_j|^2$). The choice of operator was tuned as additional method hyperparameters for SiNE and SIGNet. Logistic Regression with 5 fold cross validation of the regularization parameter was then used as binary classifier. The model was trained with the edge embeddings corresponding to edges $\{i,j\} \in E_{train}$ and labels $\{-1,1\}$. Predictions for the edges $\{i,j\} \in E_{test}$ were then computed from their corresponding edge embeddings.

\begin{table*}[h!]
	\caption{Sign prediction AUC for all networks. Best performing method per dataset is highlighted in bold.}
	\label{tab:result-auc}
	\begin{tabular}{l|cccccc|c}
		\toprule
		Methods & Slashdot(a) & Slashdot(b) & Epinions & Wiki-rfa & Bitcoin-$\alpha$ & Bitcoin-otc & Avg. AUC Rank \\ 
		\midrule
		SiNE	&0.850$\pm$0.002	&0.856$\pm$0.002	&0.898$\pm$0.001	&0.816$\pm$0.002	&0.835$\pm$0.012	&0.857$\pm$0.005	&8 \\ 
		nSNE	&0.895$\pm$0.001	&0.894$\pm$0.002	&0.950$\pm$0.001	&0.879$\pm$0.002	&0.810$\pm$0.024	&0.868$\pm$0.010	&4.5 \\ 
		lSNE	&0.886$\pm$0.002	&0.893$\pm$0.002	&0.941$\pm$0.002	&0.872$\pm$0.001	&0.854$\pm$0.022	&0.907$\pm$0.005	&6 \\ 
		SIGNet	&0.887$\pm$0.002	&0.893$\pm$0.002	&0.943$\pm$0.001	&0.848$\pm$0.005	&0.893$\pm$0.002	&0.928$\pm$0.006	&4.5 \\ 
		Prior(pol.)	&0.879$\pm$0.001	&0.885$\pm$0.001	&0.929$\pm$0.001	&0.870$\pm$0.001	&0.858$\pm$0.014	&0.891$\pm$0.002	&7 \\ 
		Prior(pol.+tri.)	&0.888$\pm$0.001	&0.894$\pm$0.001	&0.937$\pm$0.001	&0.875$\pm$0.001	&0.874$\pm$0.009	&0.914$\pm$0.004	&3 \\ 
		CSNE(pol.)	&0.902$\pm$0.001	&0.908$\pm$0.001	&0.952$\pm$0.001	&0.886$\pm$0.002	&0.896$\pm$0.009	&0.930$\pm$0.002	&2 \\ 
		CSNE(pol.+tri.)	&\textbf{0.904$\pm$0.001}	&\textbf{0.909$\pm$0.001}	&\textbf{0.954$\pm$0.000}	&\textbf{0.887$\pm$0.002}	&\textbf{0.899$\pm$0.006}	&\textbf{0.936$\pm$0.002}	&1 \\ 
		\bottomrule
	\end{tabular}
\end{table*}

\ptitle{Hyperparameters} Throughout our evaluation, we set number of dimensions $|d| = 20$ for all methods. For CSNE we fixed the spread parameters $\sigma_1=1$ and $\sigma_2=2$. We ran SiNE, nSNE, lSNE and CSNE for 500 iterations. This value was empirically found to provide best results in preliminary experiments. Fitting a MaxEnt prior is a convex problem for which we use second order information, therefore, we limited the number of iterations in this case to 20. We performed method hyperparameter tuning on a validation set obtained by further splitting $E_{train}$ in 80\% training and 20\% validation. The specific method hyperparameters we tuned are as follows. For SiNE, we tuned $\delta = \delta_0 \in \{0.5,1\}$ and the edge embedding operator. For SIGNet we only tuned the edge embedding operator. For nSNE and lSNE, we varied $\lambda \in \{5e-5, 2.5e-5, 1e-5\}$ and $\beta \in \{0.5, 0.05, 0.005, 0\}$. Finally, for CSNE, including the MaxEnt priors, we did not tune any hyperparameters. 


\ptitle{Evaluation measure} We evaluated the methods in terms of Area Under the Curve for the Receiver Operating Characteristic (AUC-ROC). This metric is popular for binary classification tasks and well suited for prediction in the case of class imbalance. Let TP, TN, FP and FN be the elements of a confusion matrix. Then, we can compute the true positive rate as $TPR = \frac{TP}{TP+FN}$ and the false positive rate as $FPR = \frac{FP}{FP+TN}$. The AUC is then the area under the ROC curve created by plotting the true positive rate (TPR) against the false positive rate (FPR) at various thresholds.

\subsection{Reproducibility\label{ssec_rep}}
In order to enhance the reproducibility of our experimental evaluation we used the \texttt{EvalNE} toolbox \cite{mara2019evalne}. This Python toolbox aims to simplify and standardize the evaluation of network embedding methods on various downstream tasks. \texttt{EvalNE} uses configuration files that detail the evaluation pipeline. These config files together with an implementation of CSNE and instructions for replicating our experiments are available online~\footnote{\url{https://bitbucket.org/ghentdatascience/csne-public}}. This files, together with the datasets and methods reported in Secs.~\ref{ssec_data} and \ref{ssec_methods} allow full reproduction of our experiments.


\section{Experimental Results}\label{sec_results}
In this section we present and discuss quantitative and qualitative experimental results. Quantitative results, on one hand, are shown for the task of sign prediction. A qualitative evaluation, on the other, is performed through visualization of the signed embeddings learned by CSNE on a small network representing relations between characters in the Harry Potter novels.  

\subsection{Sign Prediction}\label{ssec_spRes}

We start in Table~\ref{tab:result-auc} by presenting the AUC scores for each method on all evaluated datasets. For CSNE we present the results using two MaxEnt priors with different sets of constraints, i.e. node polarity only (pol.) and node polarity combined with balanced/unbalanced triangle counts (pol.+tri.). These MaxEnt priors, as discussed in Sec.~\ref{ssec_maxent}, can be independently used for sign prediction. Therefore, we also include their performance in Table~\ref{tab:result-auc} as \textit{Prior(pol.)} and \textit{Prior(pol.+tri.)}, respectively.

Our results showcase the superior performance of CSNE over the baseline methods on all datasets. The largest difference in AUC to the best performing baseline (2.7\%) can be seen for Slashdot(b) while the lowest (0.4\%) can be found for Epinions. Additionally, CSNE exhibits a more consistent performance on different networks as compared to other baselines. SIGNet, the best performing method on Bitcoin-$\alpha$, performs significantly worse than CSNE and the nSNE and lSNE baselines on Wikipedia. Similarly, nSNE, the best baseline on Slashdot(a), Slashdot(b) and Epinions obtains poor results on Bitcoin-$\alpha$ and Bitcoin-otc. 

The results in Table~\ref{tab:result-auc} also show that using balanced and unbalanced triangle counts as additional structural constraints to the node polarity in the MaxEnt prior, always results in improved AUC scores. This effect is most prominent when comparing \textit{Prior(pol.)} and \textit{Prior(pol.+tri.)}, where the latter obtains higher accuracy scores across the board. A similar, yet less prominent effect can be observed when these priors are used as part of CSNE. Another interesting observation from Table~\ref{tab:result-auc} is that the proposed \textit{Prior(pol.+tri.)} already provides state-of-the-art results while being much faster than other methods, as shown by our runtime experiments.

\begin{figure}[t!]
	\centering
	\includegraphics[width=0.95\linewidth]{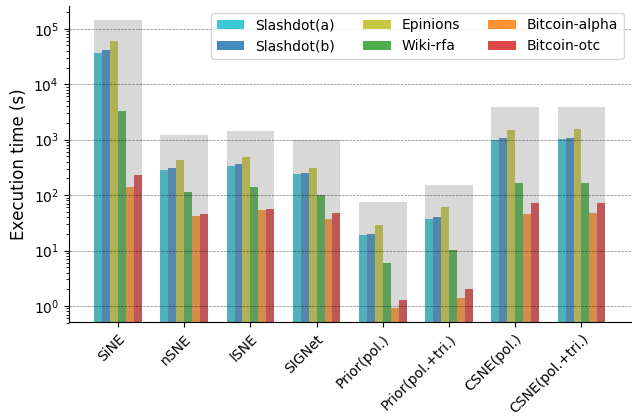}
	\caption{Execution times of all methods for sign prediction on each evaluated network. Gray boxes indicate the total runtime per method on all datasets.}
	\label{fig:exec_times}
\end{figure}

In Figure~\ref{fig:exec_times} we present the execution times in seconds (all experiments were run on a machine equipped with an Intel(R) Core i7-7700K processor and 32GB of RAM), including hyperparameter tuning, for all methods. Each colour in the figure represents a different network and the grey boxes indicate, per method, the corresponding cumulative execution times on all networks (i.e. sum of all coloured bars). An immediate observation from Figure~\ref{fig:exec_times} is that the total execution time of SiNE (grey bar in the Figure~\ref{fig:exec_times}), is approximately two orders of magnitude larger than those of other methods. The cumulative execution times on all networks of the two MaxEnt priors, on the other hand, are approximately one order of magnitude lower than those of the fastest evaluated baseline, SIGNet. We also observe that the two CSNE variants are not significantly slower, on most networks, than other baselines. 
 
\begin{figure}
	\centering
	\includegraphics[width=0.95\linewidth]{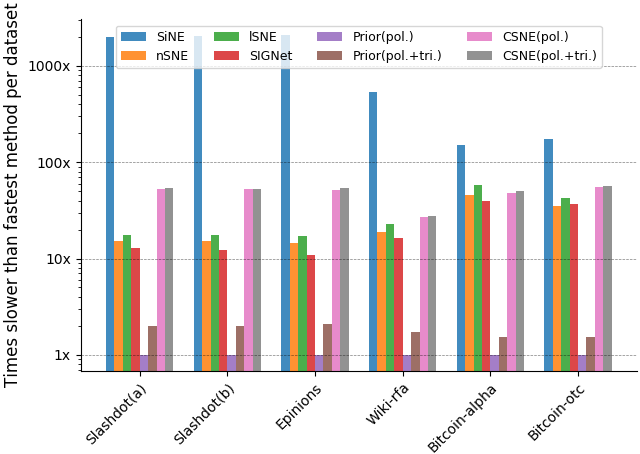}
	\caption{Relative execution times of all methods compared to the fastest approach on each dataset (lower is better).}
	\label{fig:exec_times2}
\end{figure}

In Figure ~\ref{fig:exec_times2} we group the methods by dataset and present their execution times relative to the fastest method in each case. Firstly, we observe that \textit{Prior(pol.)} is the fastest method on all networks and that \textit{Prior(pol.+tri.)} is never more than 2x slower. The remaining methods, with the exception of SiNE, are between 10x and 50x slower. For SiNE, we observe execution times that are up to 2000 times those our proposed \textit{Prior(pol.)} approach on the two Slashdot and the Epinion networks. Finally, the relative differences in execution times between the baselines and the MaxEnt priors appear to be related to the number of nodes in the network. 

\subsection{Hyperparameter Sensitivity}
We evaluated the sensitivity of the proposed CSNE method w.r.t. three hyperparameters, i.e. train and test edge size, embedding dimensionality and the spread parameter $\sigma$.

\subsubsection{Train Set Size}
\begin{figure}
	\centering
	\includegraphics[width=0.95\linewidth]{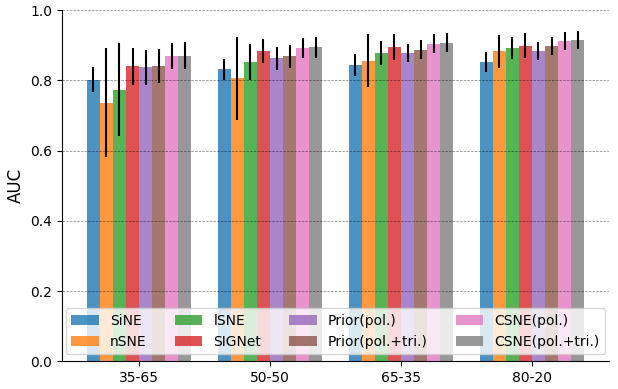}
	\caption{Average sign prediction AUC over all evaluated networks for varying sizes of $E_{train}$ and $E_{test}$. A black line over each bar denotes the 95\% confidence interval for the estimate of the average performance.}
	\label{fig:trte}
\end{figure}

First, we assessed the generalization performance of the proposed method from different amounts of initial training data. We did this by dividing the set of all graph edges $E$ in sets $E_{train}$ and $E_{test}$ of different sizes. We started by using 35\% of all edges for training and 65\% for testing. We then gradually increased the size of $E_{train}$ to 50\%, 65\% and finally 80\% while the size of $E_{test}$ scaled accordingly 50\%, 35\% and 20\%. Using the same setting as in Sec.~\ref{ssec_sp} we performed sign prediction evaluation for these different edge splits. For comparison, we also included the AUC scores of the baseline methods. The average AUC scores of each method on all the evaluated networks, are summarized in Figure~\ref{fig:trte}. The 95\% confidence intervals are also presented for each case.

The results show that the performance of lSNE, nSNE and SIGNet degrades significantly as the size of $E_{train}$ decreases. CSNE, the MaxEnt prior and SiNE, on the other hand, are more robust to changes in the size of $E_{train}$. The tight confidence intervals for these three methods also indicate a consistent performance across different datasets. For lSNE, nSNE and SIGNet these ranges are larger, especially when little training data is available.

\subsubsection{Embedding Dimensionality} 
\begin{figure}[t!]
	\centering
	\begin{minipage}{0.52\columnwidth}
		\centering
		\includegraphics[width=\linewidth]{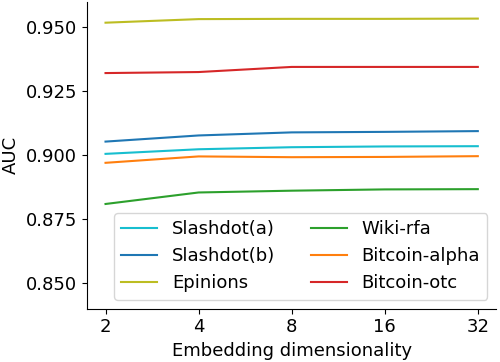}
		\subcaption{}
		\label{fig:embdim}
	\end{minipage}%
	\begin{minipage}{0.52\columnwidth}
		\centering
		\includegraphics[width=\linewidth]{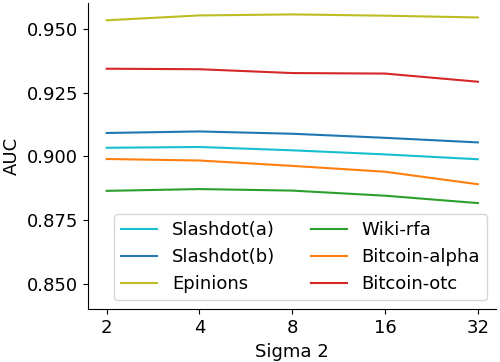}
		\subcaption{}
		\label{fig:sigma2}
	\end{minipage}
	\caption{Sign prediction AUC scores of CSNE(pol.+tri.) for (a) different embedding dimensions $d$ and (b) different values of $\sigma_2$.}
\end{figure}

Another fundamental parameter for signed network embedding methods is the size of the resulting embeddings. We studied the performance of CSNE w.r.t. the dimensionality for $d \in \{2,4,8,16,32\}$. The results, depicted in Figure~\ref{fig:embdim}, show a consistent performance of the method for all values of this parameter. Even for values as low $d=2$, the performance of out method is excellent. This indicates that CSNE can be directly used for visualization without the need to use additional tools.

\subsubsection{Spread parameter $\sigma$} The two $\sigma$ parameters introduced in Sec.~\ref{ssec_csne} determine, on one hand, the spread of the Gaussian distributions from which the distances between positively and negatively connected pairs are generated in Eq.~(\ref{eq:cond}). These parameters also control the strength of the pull and push effects in Eq.~(\ref{eq:dergrad}). Therefore, to understand the effect on performance of different values for these parameters, we conducted and additional experiment. In this experiment, we set $\sigma_1=1$, as this simply fixes the scale, and varied $\sigma_2 \in \{2,4,8,16,32\}$. The results, summarized in Figure~\ref{fig:sigma2}, show the robustness of CSNE to changes of this parameter. Only very large values of $\sigma_2$ i.e. $\sigma_2=16$ and $\sigma_2=32$ appear to have a slight effect on method performance. As such, $\sigma_1=1$ and $\sigma_2=2$ appear to be good default values in most cases. 






\subsection{Convergence Analysis}
\begin{figure}[t!]
	\centering
	\begin{minipage}{0.52\columnwidth}
		\centering
		\includegraphics[width=\linewidth]{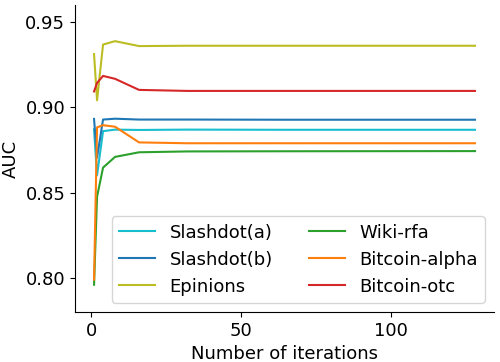}
		\subcaption{}
		\label{fig:conv_maxent}
	\end{minipage}%
	\begin{minipage}{0.52\columnwidth}
		\centering
		\includegraphics[width=\linewidth]{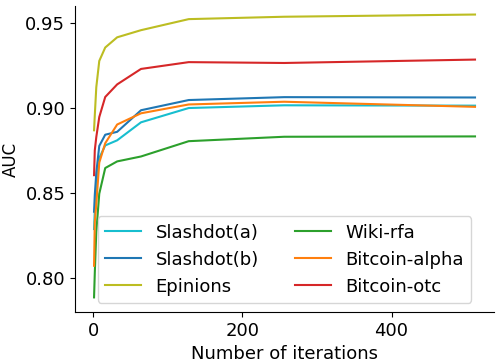}
		\subcaption{}
		\label{fig:conv_csne}
	\end{minipage}
	\caption{Convergence of (a) the MaxEnt prior and (b) CSNE.}
	\label{fig:conv}
\end{figure}

In this section we present and discuss the convergence of the MaxEnt prior and of the complete CSNE approach. In Figures~\ref{fig:conv_maxent} and \ref{fig:conv_csne} we plot, for each method, the sign prediction AUC against the iteration number on all datasets introduced in Table~\ref{tab:networks}. In both cases, we used the joint polarity and triangle count prior. Computing the MaxEnt prior amounts to optimizing a convex function as discussed in Sec.~\ref{ssec_maxent}. By leveraging second order information, convergence is achieved in less than 20 iterations for all datasets as shown in Figure~\ref{fig:conv_maxent}. In CSNE, maximizing the likelihood function is a non-convex optimization problem solved via block stochastic gradient descent. Convergence is achieved in approximately 200 iterations as shown in Figure~\ref{fig:conv_csne}.



\subsection{A Case Study: Visualization}
We also performed a qualitative evaluation of the CSNE embeddings on the popular Harry Potter network\footnote{\url{https://github.com/efekarakus/potter-network/tree/master/data}}. Nodes in the graph correspond to characters in the novels while edges denote friend or enemy relations extrapolated from the character interactions throughout the novels. We preprocessed the original directed network to obtain an undirected representation, extracted the main connected component and removed self loops. The resulting network $G$ contained $n=65$ nodes and $m=453$ edges with an average degree of $14$. 

We computed a 2-dimensional embedding $\mathbf{X}$ of $G$ using CSNE with a MaxEnt structural prior encoding node polarity and the number of balanced and unbalanced triangles. We used 100 iterations to fit the prior and another 100 iterations to learn the embeddings. We obtained an initial assessment of the embedding quality by performing sign prediction for all edges $E$ of $G$. The resulting AUC score is $0.994$.

\begin{figure}
	\centering
	\includegraphics[width=0.91\linewidth]{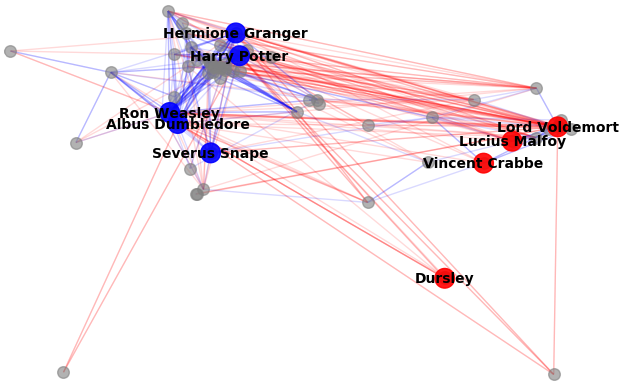}
	\caption{Plot showing the CSNE embeddings of the Harry Potter network. Enemy relationships between book characters are highlighted in red, while friendship is highlighted in blue. A subset of protagonists and antagonists are shown in blue and red, respectively.}
	\label{fig:full_d2}
\end{figure}

The embeddings learned by CSNE are presented in Figure~\ref{fig:full_d2} where blue links denote friendship relations and red links denote enemy relations. The main protagonists and antagonists of the novels are presented as blue and red circles, respectively. Two clear clusters can be identified, a larger one corresponding to the protagonists and allies and a smaller one for the antagonists. The relations within each cluster are mostly positive while between clusters are negative. To verify that, as expected, positively connected nodes are, on average, closer to each other than negatively connected ones, we used the Euclidean distance. The average Euclidean distance obtained for all positively connected pairs $\{i,j\} \in E^+$ is $0.745$ with a standard deviation of $0.525$. For the negatively connected pairs, $\{i,j\} \in E^-$, the obtained distance is $3.360\pm1.014$. This experiment shows that CSNE is able to effectively capture the structure of a signed network in dimensionalities as low as $d=2$. This showcases the potential of CSNE for signed network visualization tasks.

\section{Conclusion and Future Work}\label{sec_conclusion}
In this paper, we have presented a new probabilistic approach for learning representations of signed networks with applications to visualization and sign prediction. The proposed CSNE method solves an MLE problem which seeks the embeddings that maximize the probability of observing the signs on the edges of an input graph. Our optimization process models certain structural properties of the data as a MaxEnt prior. Particularly, this prior can capture node polarity and structural balance i.e. as counts of balanced and unbalanced triangles. Our experimental results indicate that CSNE can adequately model the specific properties of signed networks and outperforms other baselines for sign prediction. Additionally,  we have showed that the proposed MaxEnt priors can also be directly used for sign prediction, resulting in state-of-the-art AUC scores with runtimes up to 50x lower than those of other baselines. Our work opens up several avenues for further research and improvements. On one hand, more sophisticated MaxEnt priors can be designed specifically for networks with a particular structures, such as k-partiteness. Another possible line of work is to tackle the extension of CSNE to directed signed networks.


\begin{acks}
The research leading to these results has received funding from the European Research Council under the European Union's Seventh Framework Programme (FP7/2007-2013) / ERC Grant Agreement no. 615517, from the Flemish Government under the ``Onderzoeksprogramma Artificiële Intelligentie (AI) Vlaanderen'' programme, and from the FWO (project no. G091017N, G0F9816N, 3G042220).
\end{acks}
\bibliographystyle{ACM-Reference-Format}
\bibliography{bibliography}


\end{document}